\documentclass[preprint,showpacs,preprintnumbers,amsmath,amssymb]{revtex4}

\usepackage{graphicx}
\usepackage{dcolumn}
\usepackage{bm}

\begin{document}

\preprint{}
{\bf }\title{
Sensitivity of exclusive proton knockout spin observables to different Lorentz invariant
representations of the NN interaction
}
\author{B.~I.~S.~van der Ventel$^{1}$ and G.~C.~Hillhouse$^{1,2}$}  
\affiliation{$^{1}$Department of Physics, University 
of Stellenbosch, Private Bag X1, Matieland 7602, South Africa\\
$^{2}$Research Center for Nuclear Physics, Osaka
University, Ibaraki, Osaka 567-0047, Japan}
\date{\today}

\begin{abstract} 
Within the framework of the relativistic plane wave impulse approximation, 
we study the observable consequences of employing a complete Lorentz 
invariant representation of the NN scattering matrix in terms of 44 
independent amplitudes, as opposed to the previously-employed, but 
ambiguous, five-term Lorentz invariant parametrization of the NN 
scattering matrix, for the prediction of complete sets of exclusive 
($\vec{p},2 \vec{p}\,$) polarization transfer observables.  Two 
kinematic conditions are considered, namely proton knockout from 
the $3s_{1/2}$ state of $^{208}$Pb at an incident energy of 202~MeV 
for coplanar scattering angles ($28.0^{\circ}, -54.6^{\circ}$), as 
well as an incident energy of 392~MeV for the angle pair 
($32.5^{\circ}, -80.0^{\circ}$). The results indicate that certain spin observables
are ideal for discriminating between the two representations. 
\end{abstract}


\maketitle

\section{\label{sec:intro}Introduction}
The representation of the nucleon-nucleon (NN) scattering matrix, within 
the context of an appropriate dynamical framework is crucial for the
description of nuclear reactions and nuclear structure. 

Recently, we demonstrated that a model based on the relativistic distorted 
wave impulse approximation (DWIA) provides an almost perfect description 
of exclusive ($\vec{p}, 2 p$) analyzing power data, whereas corresponding 
nonrelativistic Schr\"{o}dinger-equation-based predictions completely fail 
\cite{Ne02,Hi03a,Hi03b}. For both dynamical models, however, the comparison 
of theoretical predictions to unpolarized cross section data yields similar 
spectroscopic factors which are also in good agreement with those extracted 
from ($e,e'p$) studies. The above results highlight the important role 
that spin observables (such as the analyzing power) play, as opposed 
to unpolarized cross sections, in effectively discriminating between 
different dynamical effects and, at the same time, also pointing to the Dirac 
equation as the preferred equation of motion. 

However, before claiming the latter statement with absolute certainty, it is 
necessary to subject our relativistic DWIA models to additional tests, such 
as comparing model predictions to additional spin observable data, other than 
the commonly measured analyzing power. In particular, it is important to 
eliminate obvious ambiguities associated with the choice of representation 
for the NN scattering matrix. Essentially, the problem is related to the 
direct application of the free on-shell NN scattering matrix for the 
description of nucleons scattering from nuclei: the external nucleons 
partaking in free on-shell NN scattering are represented by free 
positive-energy Dirac spinors, whereas for scattering from nuclei the 
scattering wave functions are linear combinations of both positive 
and negative- energy Dirac spinors [see Sec.~(\ref{sec:results})]. To date, 
all applications of Dirac relativity to describe exclusive ($p,2 p$) 
reactions have adopted the so-called IA1 parametrization, whereby the free 
NN scattering matrix is parametrized in terms of five Lorentz (scalar, 
pseudoscalar, vector, axial-vector, and tensor) invariants which are 
consistent with parity and time-reversal invariance as well as charge 
symmetry. These invariant amplitudes are obtained by fitting to free 
NN scattering data.  There are, however, an infinite number of 
five-term representations with the same on-shell matrix elements for free 
NN scattering, and which also respect the above symmetries \cite{Go57,Br76}. 
Hence, free on-shell NN scattering data cannot distinguish between different five-term 
representations. However, for applications to nuclear reactions, different five-term 
representations result in drastically different observables, thereby clouding
physical interpretation of the data. In Refs.~\cite{Hi94,Hi95,Hi98} we demonstrated 
the limitations of the IA1 representation for applications to inclusive quasielastic 
proton-nucleus scattering: for example, one type of five-term representation 
describes ($\vec{p},\vec{p}\,'$) data, whereas a different representation is required 
for ($\vec{p},\vec{n}$) scattering. 

The ambiguities associated with the IA1 representation can be eliminated by employing 
the more appropriate, but more complicated, IA2 representation developed by Tjon and 
Wallace \cite{Tj85a,Tj85b,Tj87a,Tj87b}, whereby the NN scattering matrix is expanded 
in terms of a complete set of 44 independent invariant amplitudes consistent with 
the above-mentioned symmetries. It follows therefore that IA1 neglects 39 additional 
amplitudes that should appear on the grounds of very general symmetry principles.
The aim of this paper is to apply the IA2 
representation, for the first time, to exclusive proton knockout reactions 
and to compare results to corresponding IA1 predictions of complete sets of 
spin observables.

We have already studied the observable consequences of employing the IA2, 
rather than the IA1 representation, for describing inclusive quasielastic 
proton-nucleus scattering \cite{Va99,Va00}. The aim of the latter paper was to 
identify spin observables which are sensitive enough to extract information 
regarding the modification of the properties of the strong nuclear force by 
nuclear matter. In particular, we demonstrated that the IA1 representation severely
overestimates the importance of nuclear medium modifications on the NN interaction, 
whereas application of the IA2 representation suggests that quasielastic spin observables 
are insensitive to these effects. This emphasizes the critical role played by the
representation of the NN scattering matrix in giving a correct interpretation of
the results.

Although we have already demonstrated that a quantitative description of 
($\vec{p},2 \vec{p}\,$) spin observable data requires the inclusion of nuclear 
medium effects on the scattering wave functions \cite{Hi03a}, in this paper 
we consider a relativistic plane wave approximation, thus neglecting the 
role of distorting optical potentials on the scattering wave functions. 
This simplification will allow us to uniquely focus on the effect of the 
different representations employed for $\hat{F}$.
Two kinematic conditions are considered, namely proton knockout from the $3s_{1/2}$ 
state of $^{208}$Pb at an incident energy of 202~MeV for coplanar scattering angles 
$(\theta_{a'},\theta_{b}) \, = \, (28.0^{\circ}, -54.6^{\circ})$, 
as well as an incident energy of 392~MeV for coplanar scattering angles
$(\theta_{a'},\theta_{b}) \, = \, (32.5^{\circ}, -80.0^{\circ})$.
The reaction kinematics at 202~MeV are chosen to correspond to recent measurements by 
Neveling {\it et al.} \cite{Ne02} at iThemba Laboratory for Accelerator Based Sciences 
(Faure, South Africa), and the kinematics at 392~MeV correspond to present and future 
experimental programs at the Research Center for Nuclear Physics in Osaka, Japan \cite{No03}.
In addition, the above kinematics are also chosen so as to minimize complications associated 
with the inclusion of recoil corrections in the Dirac equation \cite{Co93a,Ma93a}, while 
still maintaining the validity of the impulse approximation. 

In Sec.~(\ref{sec:rpwia}), we present the formalism for the relativistic plane wave impulse
approximation. Thereafter, in Sec.~(\ref{sec:nn-interaction}), we derive expressions for 
the relativistic scattering matrix element based on both IA1 and IA2 representations of 
the NN interaction. The expressions for calculating complete sets of spin observables 
are presented in Sec.~(\ref{sec:pto}). Results are given in Sec.~(\ref{sec:results}), 
and we summarize and draw conclusions in Sec.~(\ref{sec:conclusions}).

\section{\label{sec:rpwia}Relativistic Plane Wave Impulse Approximation}
\begin{figure}
\includegraphics{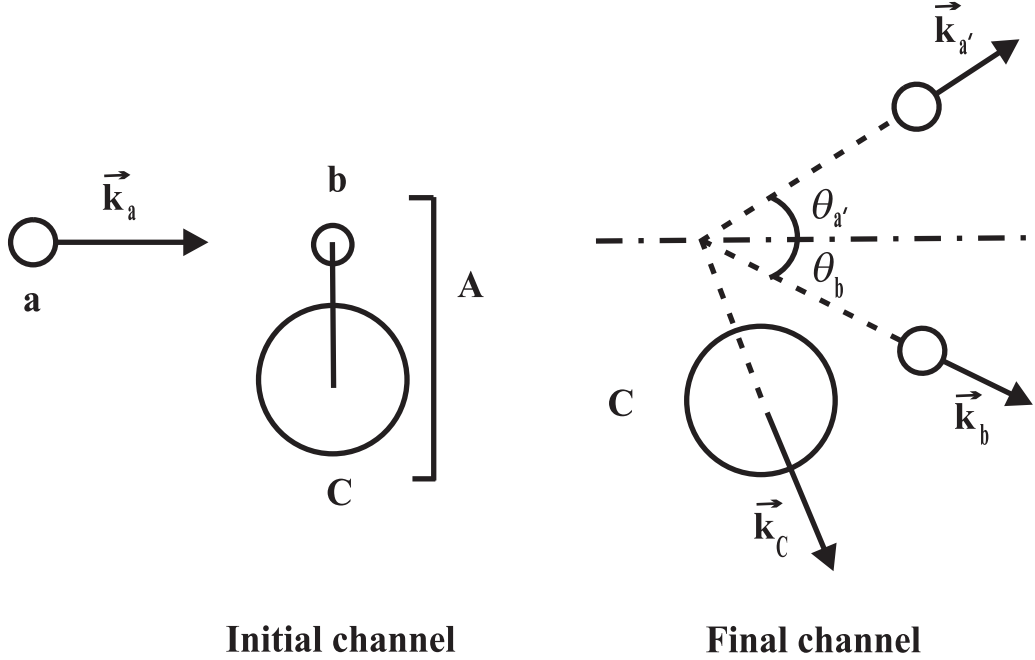}
\caption{Schematic representation for the 
coplanar $(p,2p)$ reaction of interest.}
\label{fig-p2pgeometry}
\end{figure}

Consider an exclusive $(p,2p)$ reaction, written as $A(a,a'b)C$ for notational 
purposes and depicted schematically in Fig.~(\ref{fig-p2pgeometry}), whereby an 
incident proton $a$ knocks out a bound proton $b$ from a specific orbital in the 
target nucleus $A$, resulting in three particles in the final state, namely the 
recoil residual nucleus $C$ and two outgoing protons, $a'$ and $b$, which are 
detected in coincidence at coplanar laboratory scattering angles (on opposite sides 
of the incident beam), $\theta_{a'}$ and $\theta_{b}$, respectively. 
All kinematic quantities are completely determined by specifying
the rest masses, $m_{i}$, of particles, where 
$i$ = ($a$,$A$, $a'$, $b$, $C$), the laboratory kinetic energy 
$\mbox{T}_{a}$ of incident particle $a$, the laboratory kinetic energy 
$\mbox{T}_{a'}$ of scattered particle $a'$, the laboratory scattering angles 
$\theta_{a'}$ and $\theta_{b'}$, and also the binding energy 
of the proton that is to be knocked out of the target nucleus $A$.
In this paper we employ the conventions of Bjorken and Drell \cite{Bj64} and,
unless otherwise stated, all kinematic quantities are expressed in natural
units (i.e., $\hbar = c =1$).
For a zero range approximation to the NN interaction, the relativistic transition matrix 
element associated with Fig.~(\ref{fig-p2pgeometry}) is given by \cite{Hi03a,Hi03b}:
\begin{eqnarray}
\hspace{-9mm}
\label{eq_1}
 T_{LJM_{J}} (s_{a}, s_{a'}, s_{b})
 & = &
 \int d^{3} \vec{x}
 \left[
       \bar{\psi}^{(-)} (\vec{x}, \vec{k}_{a'}, s_{a'} \, ) \otimes
       \bar{\psi}^{(-)} (\vec{x}, \vec{k}_{b}, s_{b}) 
 \right] \hat{F}
 \left[
       \psi^{(+)} (\vec{x}, \vec{k}_{a}, s_{a}) \otimes \phi_{LJ M_{J}} (\vec{x})
 \right]
\end{eqnarray}
where $\otimes$ denotes the Kronecker product. The four-component
scattering wave functions, $\psi(\vec{x}, \vec{k}_{i},s_{i})$, 
are solutions to the fixed-energy Dirac scattering equations: 
$\psi^{(+)}(\vec{x}, \vec{k}_{a},s_{a})$ is the relativistic scattering 
wave function associated with the incident particle, $a$, with outgoing boundary 
conditions [indicated by the superscript $(+)$], where $\vec{k}_{a}$ 
is the momentum of particle $a$ in the laboratory frame, 
and $s_{a}$ is the spin projection thereof with respect to 
$\vec{k}_{a}$ as the $\hat{z}$-quantization axis; 
$\bar{\psi}^{(-)}(\vec{x}, \vec{k}_{j},s_{j})$
is the adjoint relativistic scattering wave function for particle $j$ 
[ $j$ = ($a',b$)] with incoming boundary conditions [indicated by the 
superscript $(-)$], where $\vec{k}_{j}$ is the momentum of particle
$j$ in the laboratory frame, and $s_{j}$ is the spin projection 
thereof with respect to $\vec{k}_{j}$ as the $\hat{z}$-quantization axis.
The boundstate proton wave function, $\phi^{B}_{L J M_J}(\vec{x}\, )$, 
labeled by single-particle quantum numbers $L$, $J$, and $M_{J}$, is given by:
\begin{eqnarray}
\label{eq_2}
 \phi_{LJ M_{J}} (\vec{x} \,)
 & = &
 \frac{1}{x} \, \sum_{s_{B}} \,
 \left(
  \begin{array}{c}
   u_{LJ}(x) \, \langle L, M_{J}-s_{B}, \frac{1}{2}, s_{B} \, | J, M_{J} \, \rangle \,
   Y_{L,M_{J}-s_{B}} (\hat{x}) \, \chi_{s_{B}}
   \\
   iw_{LJ}(x) \, \langle 2J-L, M_{J}-s_{B}, \frac{1}{2},s_{B} \, | \, J, M_{J} \, \rangle \,
   Y_{2J-L,M_{J}-s_{B}} (\hat{x}) \, \chi_{s_{B}}
  \end{array}
 \right)
\end{eqnarray}
where the brackets $<\ >$ and $Y$ denote the usual Clebsch-Gordan 
coefficients and spherical harmonics, respectively, 
$s_{B} = \pm \frac{1}{2}$, and
\begin{eqnarray}
\label{eq_5c}
\chi_{s_{B} \, = \, -\frac{1}{2}} \, = \,
\left(
      \begin{array}{c}
      0
      \\
      1
      \end{array}
\right)
& \qquad \mbox{and} \qquad &
\chi_{s_{B} \, = \, \frac{1}{2}} \, = \,
\left(
      \begin{array}{c}
      1
      \\
      0
      \end{array}
\right)\,.
\end{eqnarray}
The upper- and lower-component radial wave functions, $u_{LJ}(x)$ and $w_{LJ}(x)$ 
respectively, are obtained via selfconsistent solution of the Dirac-Hartree field 
equations of quantum hadrodynamics \cite{Ho81}. 

$\hat{F}$ denotes the relativistic NN scattering matrix. 
In this paper we will consider two different representations of $\hat{F}$ 
[see Sec.~(\ref{sec:nn-interaction})] and, in particular, study the 
sensitivity of exclusive ($\vec{p},2 \vec{p}\,$) polarization transfer 
observables to these representations. In a previous paper we 
demonstrated the importance of including distorting optical 
potentials on the incident and outgoing scattering wave functions
for a correct description of ($\vec{p},2p$) analyzing powers \cite{Hi03a}. 
However, in order to simplify the present analysis, we consider a relativistic
plane wave model, whereby all distorting optical potentials are set equal 
to zero in the Dirac equation. This approximation will allow us to uniquely 
focus on the effect of the different representations employed for $\hat{F}$. 
Secondly, plane wave calculations always form a baseline against which full 
distorted wave calculations must be tested. With these caveats in mind we 
now proceed to derive an expression for $T_{LJ M_{J}}$ based on the 
relativistic plane wave approximation.

The scattering solutions to the free Dirac equation are given by
\begin{eqnarray}
\label{eq_3}
 \psi^{(+)} (\vec{x}, \vec{k}_{a}, s_{a})
 & = &
 e^{i \vec{k}_{a} \cdot \vec{x}} \, U(\vec{k}_{a}, s_{a}),\nonumber\\
 \bar{\psi}^{(-)} (\vec{x}, \vec{k}_{a'}, s_{a'} \, )
 & = &
 e^{-i \vec{k}_{a'} \cdot \vec{x}} \, \overline{U}(\vec{k}_{a'}, s_{a'} \,),\nonumber\\
 \bar{\psi}^{(-)} (\vec{x}, \vec{k}_{b}, s_{b} \, )
 & = &
 e^{-i \vec{k}_{b} \cdot \vec{x}} \, \overline{U}(\vec{k}_{b}, s_{b} \,)
\label{eq_pwspinors}
\end{eqnarray}
where the Dirac spinor
\begin{eqnarray}
\label{eq_5a}
 U(\vec{k}_{i}, s_{i})
 & = &
 \left[
       \frac{E_i + m_i}{2 m_i}
 \right]^{\frac{1}{2}} \,
 \left(
       \begin{array}{c}
        \chi_{s_{i}}
	\\
	\displaystyle \frac{\vec{\sigma} \cdot \vec{k_{i}}}{E_i + m_{i}} \, \chi_{s_{i}}
       \end{array}
 \right)
\end{eqnarray}
is normalized such that
\begin{eqnarray}
\overline{U}(\vec{k}_{i}, s_i) \, U(\vec{k}_{i}, s_i)\ =\ 1\,.
\end{eqnarray}
$\chi_{s_{i}}$ refers to the usual 2-component Pauli spinors defined in Eq.~(\ref{eq_5c}),
$m_i$ denotes the rest mass of particle $i$, and $E_{i}\ =\ \sqrt{\vec{k}^2_i + m^2_i}$.
Substitution of Eqs.~(\ref{eq_pwspinors}) and (\ref{eq_5a}) into Eq.~(\ref{eq_1}), 
results in the following expression for the transition matrix element:
\begin{eqnarray}
\label{eq_6}
 T_{LJ M_{J}} (s_{a}, s_{a'}, s_{b})
 & = &
 \left[
       \overline{U}(\vec{k}_{a'},s_{a'} \,) \otimes \overline{U} (\vec{k}_{b},s_{b})
 \right] \, \hat{F} \,
 \left[
       U(\vec{k}_{a},s_{a}) \otimes \phi_{LJ M_{J}} (\vec{K} \,)
 \right]
\end{eqnarray}
with
\begin{eqnarray}
\label{eq_7}
 \phi_{LJ M_{J}} (\vec{K} \, ) \, = \, \phi_{LJ M_{J}} (-\vec{k}_{C} \,)
 & = &
 \int \, d^{3} \vec{x} \, \, e^{-i \vec{K} \cdot \vec{x}} \, 
 \phi_{LJ M_{J}} (\vec{x} \, ),
\end{eqnarray}
where $\vec{k}_{C}$ is the recoil three-momentum of the residual nucleus given by
\begin{eqnarray}
\label{eq_8}
 \vec{k}_{C} & = & -\vec{K} \, = \, -(\vec{k}_{a'} + \vec{k}_{b} - \vec{k}_{a}).
\end{eqnarray}
Equation~(\ref{eq_6}) may be interpreted as the transition matrix element for a two-body scattering 
process in which the initial proton is bound. Combining Eqs.~(\ref{eq_2}) and (\ref{eq_7}) yields:
\begin{eqnarray}
\label{eq_9}
 \phi_{LJ M_{J}} (\vec{K} \, ) \, = \, \phi_{LJ M_{J}} (-\vec{k}_{C} \,)
 & = &
 \left(
  \begin{array}{c}
       4 \pi i^{L} {\cal Y}_{LJ M_{J}} (\theta_{k_{C}}, \phi_{k_{C}}) \, u_{LJ} (k_{C})
       \\
       4 \pi i^{2J-L+1} {\cal Y}_{2J-L+1,J M_{J}} (\theta_{k_{C}}, \phi_{k_{C}}) \, 
       w_{2J-L,J} (k_{C})
  \end{array}
 \right)
\end{eqnarray}
with
\begin{eqnarray}
\label{eq_10}
 u_{LJ} (k_{C}) & = & \int_{0}^{\infty} \, dx \ x\ j_{L} (k_{C} x)\ u_{LJ} (x),
 \\
\label{eq_11} 
 w_{2J-L,J} (k_{C}) & = & \int_{0}^{\infty} \, dx \ x\ j_{2J-L} (k_{C} x)\ w_{LJ} (x),\\
\label{eq_12}
 {\cal Y}_{LJ \mu} (\theta, \phi)
 & = &
 \sum_{s_{z}'} \ \langle L, \frac{1}{2}, \mu-s_{z}', s_{z}' \, | J \mu \rangle \
 Y_{L, \mu-s_{z}'} (\theta, \phi) \ \chi_{s_{z}'},
\end{eqnarray}
where $j_{L}(k_{C} x)$ denotes the usual spherical Bessel functions.

\section{\label{sec:nn-interaction}
Lorentz invariant representations of the NN interaction}
The relativistic scattering matrix $\hat{F}$ is one of the principal components in the
calculation of the transition matrix element. Since we are assuming the impulse approximation to
be valid, we employ the free NN interaction for $\hat{F}$. The purpose of this investigation is to
study how sensitive the polarization transfer observables [to be defined in Sec.~(\ref{sec:pto})] 
are to two different representations of $\hat{F}$. The first form of $\hat{F}$ which we will employ 
is known as the IA1 representation and is a parametrization of the scattering matrix in 
terms of five complex amplitudes \cite{Mc83a}:
\begin{eqnarray}
\label{eq_13}
 \hat{F} & = & \sum_{L=S}^{T} \, F_{L} \, \left( \lambda^{L} \otimes \lambda_{L} \right)
\end{eqnarray}
where
\begin{eqnarray}
\label{eq_14}
 \lambda^{L} & \in & \{ I_{4}, \gamma^{5}, \gamma^{\mu}, \gamma^{5} \gamma^{\mu}, \sigma^{\mu \nu} \}.
\end{eqnarray}
The IA1 representation of $\hat{F}$ has been used in relativistic descriptions of elastic- 
\cite{Sh83,Mc83a,Mc83b,Ho85} and inelastic \cite{Ro84,Sh84,Ro87} scattering, 
as well as inclusive quasielastic \cite{Ho86,Ho88,Hi94,Hi95,Hi98} proton-nucleus 
scattering. This five-term representation is consistent with parity
and time-reversal invariance as well as charge symmetry.
However, other five-term representations which respect the above-mentioned 
symmetries are also possible. For example, there are the GNO
invariants \cite{Go57} as well as the perturbative invariants \cite{Br76}.
The invariant amplitudes in each of the representations of $ \hat{F} $
are connected via matrix relations given in Ref. \cite{Br76} and are
obtained by fitting to free scattering data \cite{Tj85a}. Physical NN scattering
data therefore completely determine the amplitudes in a five-term
representation of $ \hat{F} $. A priori there is no reason why one
five-term representation should be chosen above another. The IA1 representation
form is very convenient since its amplitudes are free of kinematic singularities 
at $ \theta \, = \, 0 $ and $ \theta \, = \, \pi $ ($ \theta $ is the centre-of-mass 
scattering angle) and the one-meson exchange contributions are naturally written in 
terms of Fermi covariants \cite{Go60}.

However, five-term representations are ambiguous, since the application 
of different parametrizations to describe nucleons scattering from nuclei, 
as opposed to free on-shell NN scattering, gives different predictions for
the same observables. Several authors \cite{Tj85a,Pi86,Tj87a} have addressed the 
problem of eliminating the ambiguities associated with the IA1 representation
by determining a general Lorentz invariant representation of $ \hat{F} $. The 
formalism of J.~A.~Tjon and S.~J.~Wallace (referred to as the IA2 representation 
of $ \hat{F} $) will be used in the present study: this is a general and complete 
Lorentz invariant representation, whereby the NN scattering matrix is expanded in 
terms of a complete set of 44 independent invariant amplitudes consistent with 
the above-mentioned symmetries. Five of the 44 amplitudes are determined from 
free NN scattering data, and are therefore identical to the amplitudes associated 
with IA1 representation. The remaining 39 amplitudes are obtained via solution of 
the Bethe-Salpeter equation employing a one-boson exchange model for the NN interaction.
 
The IA2 representation has the attractive feature that it reduces 
to the IA1 representation explicitly as a special case.
This representation has been successfully applied to describe elastic 
\cite{Tj87b,Sa98} and inelastic proton-nucleus \cite{Ke94,Sa98,Va99,Va00} scattering.
In the IA2 representation the NN scattering matrix is given by \cite{Tj87b}:
\begin{eqnarray}
 \label{eq_19}
   \hat{F} & = &
   \sum_{\rho_{1}\, \rho'_{1} \,  \rho_{2}\, \rho'_{2}} \,
   \sum_{n = 1}^{13} \, F_{n}^{ \{ \rho_{1}\, \rho'_{1}\, ;\, \rho_{2}\, \rho'_{2}\} } \,
   [
     \Lambda_{\rho'_{1}} ( \vec{p} \, ' _{1}; m ) \otimes
     \Lambda_{\rho'_{2}} ( \vec{p} \, ' _{2}; m )
   ] \, K_{n} \,
   \nonumber
   \\
   & &
   [
    \Lambda_{\rho_{1}} ( \vec{p}_{1}; m ) \otimes
    \Lambda_{\rho_{2}} ( \vec{p}_{2}; m )
   ]
\end{eqnarray}
for a general two-body scattering process with three momenta 
($\vec{p}_{1}, \vec{p}_{2}\,$) and ($\vec{p} \, ' _{1} \, , \vec{p} \, ' _{2}\,$) 
in the initial and final channels, respectively.
Here we take $m_{1} = m_{1'} = m_{2} = m_{2'} = m$, where $m$ is the free nucleon mass.
In Eq.~(\ref{eq_19}), the invariant amplitudes for each rho-spin
sector are denoted by $F_{n}^{\{ \rho \}}$ ($n$ = 1 -- 13), where
$\{ \rho \}\ \equiv\ \{\rho_{1}\, \rho_{1}'\,;\, \rho_{2}\, \rho'_{2} \}$, 
and $\rho = \pm$; $\Lambda_{\rho} (\vec{p}, M)$ represents an energy 
projection operator defined as
\begin{eqnarray}
 \label{rho_spin_projection_operator}
  \Lambda_{\rho} ( \vec{p}, m ) 
  & \; \; = \; \; &
  \frac{\rho ( E \gamma^{0} \, - \, \vec{p} \cdot \vec{\gamma}) + m }{2 m},
\end{eqnarray}
where $ E \, = \, \sqrt{\vec{p}^{\, \, 2} \, + \, m^{2}} $, and 
the $K_{n}$'s are kinematic covariants constructed from the Dirac matrices \cite{Tj87b}.
Using Eqs.~(\ref{eq_13}) and (\ref{eq_19}) one can now write down expressions
for $T_{LJ M_{J}}$ in Eq.~(\ref{eq_6}) based on both IA1 and IA2 representations 
of the NN scattering matrix. For the IA1 representation one obtains
\begin{eqnarray}
\nonumber
 T_{LJ M_{J}} (s_{a}, s_{a'}, s_{b})
 & = &
 \sum_{L=S}^{T} \, F_{L} \,
 \left[
       \overline{U} (\vec{k}_{a'}, s_{a'} \, ) \otimes \overline{U} (\vec{k}_{b}, s_{b})
 \right] \, \left( \lambda^{L} \otimes \lambda_{L} \right)
 \\
 & &
 \label{eq_20}
 \left[
       U (\vec{k}_{a}, s_{a}) \otimes \phi_{LJ M_{J}} (\vec{K} \,)
 \right],
\end{eqnarray}
and for the IA2 representation
\begin{eqnarray}
\nonumber
 T_{LJ M_{J}} (s_{a}, s_{a'}, s_{b})
 & = &
 \sum_{n=1}^{13} \, \sum_{\{ \rho \}} \, F_{n}^{\{ \rho \}} \,
 \left[
       \overline{U} (\vec{k}_{a'}, s_{a'} \, ) \otimes \overline{U} (\vec{k}_{b}, s_{b})
 \right] \, 
 \left[
       \Lambda_{\rho_{a'}} (\vec{k}_{a'}, M) \otimes \Lambda_{\rho_{b}} (\vec{k}_{b}, m)
 \right]
 \\
 & &
\label{eq_21}
 \, K_{n} \, \left[
       \Lambda_{\rho_{a}} (\vec{k}_{a}, m) \otimes \Lambda_{\rho_{2}} (\vec{K}, m)
 \right] \,
 \left[
       U(\vec{k}_{a}, s_{a}) \otimes \phi_{LJ M_{J}} (\vec{K} \, )
 \right]\,.
\end{eqnarray}
Employing well-known identities for the energy projection operators \cite{Bj64},
results in Eq.~(\ref{eq_21}) simplifying to 
\begin{eqnarray}
\nonumber
 T_{LJ M_{J}} (s_{a}, s_{a'}, s_{b})
 & = &
 \sum_{n=1}^{13} \, \sum_{\rho_{2} = \pm} \, F_{n}^{++, \rho_{2} +} \,
 \left[
       \overline{U} (\vec{k}_{a'}, s_{a'} \, ) \otimes \overline{U} (\vec{k}_{b}, s_{b})
 \right] \, K_{n} \,
 \\
 & &
 \label{eq_22}
 \left[
       U(\vec{k}_{a}, s_{a}) \otimes \Lambda_{\rho_{2}} 
(\vec{K}, m) \, \phi_{LJ M_{J}} (\vec{K} \,)
 \right].
\end{eqnarray}
Note the presence of the energy projection operator acting on the
$\phi_{LJ M_{J}} (\vec{K})$ in Eq.~(\ref{eq_22}) as compared 
to the absence thereof in Eq.~(\ref{eq_20}). 
From Eq.~(\ref{eq_22}) we see that only subclasses 
$F^{\{+\, +\, ; \, +\, + \}}$ and
$F^{\{+\, +\, ; \, -\, + \}}$
contribute to the calculation of the transition matrix element for the IA2 
representation. This simplification occurs only because we are using plane wave 
Dirac spinors for the projectile and ejectile nucleons. The boundstate 
wave function is therefore the only remaining
spinor that has negative-energy content [see Eq.~(\ref{eq_31})]. If we had chosen 
distorted waves for the projectile and ejectile nucleons, then all 16 subclasses 
of $\hat{F}$ would have contributed. Note that the amplitudes in subclass 
$F^{\{+\, +\, ; \, +\, + \}}$ 
are determined by free NN scattering data. The amplitudes
for the other subclasses are determined from a dynamical model 
\cite{Va83,Va84}, but they can only make a 
contribution if the spinor contains a negative-energy component. 

\section{\label{sec:pto}Polarization transfer observables}
The spin observables of interest are denoted by $\mbox{D}_{\rm{i' j}}$ 
and are related to the probability
that an incident beam of particles $a$, with spin-polarization $j$, induces a 
spin-polarization $i'$ for the scattered beam of particles $a'$:
the subscript $j = (0,\rm{l,n,s})$ is used to specify the polarization of the incident
beam $a$ along any of the orthogonal directions:
\begin{eqnarray}
\hat{\rm{l}}\ =\ \hat{z}\ =\ \hat{k}_{a} \nonumber\\
\hat{\rm{n}}\ =\ \hat{y}\ =\ \hat{k}_{a} \times \hat{k}_{a'} \nonumber\\
\hat{\rm{s}}\ =\ \hat{x}\ =\ \hat{\rm{n}} \times \hat{\rm{l}}\,,
\label{e-lns}
\end{eqnarray}
and the subscript $i' = (0,\rm{l}',\rm{n}',\rm{s}')$ denotes the polarization of the scattered 
beam $a'$ along any of the orthogonal directions:
\begin{eqnarray}
\hat{\rm{l}}'\ =\ \hat{z}'\ =\ \hat{k}_{a'} \nonumber\\
\hat{\rm{n}}'\ =\  \hat{\rm{n}}\ =\ \hat{y} \nonumber\\
\hat{\rm{s}}'\ =\ \hat{x}'\ =\ \hat{\rm{n}} \times \hat{\rm{l}}'\,.
\label{e-lpnsp}
\end{eqnarray}
With the above coordinate axes in the initial and final channels, the 
spin observables, $\mbox{D}_{\rm{i' j}}$, are defined by
\begin{eqnarray}
\mbox{D}_{\rm{i' j}}\ =\ \frac{\sum_{M_{J},s_{b}} \, \mbox{Tr}(T \sigma_{j} T^{\dagger} \sigma_{i'})}
{\sum_{M_{J}, s_{b}} \mbox{Tr}(T T^{\dagger})}\,,
\label{e-dipj}
\end{eqnarray}
where $\mbox{D}_{\rm{n 0}}=\mbox{P}$ refers to the induced polarization, 
$\mbox{D}_{\rm{0 n}} = \mbox{A}_{\rm{y}}$ denotes
the analyzing power, and the polarization transfer observables of interest are
$\mbox{D}_{\rm{n n}},\, \mbox{D}_{\rm{s' s}},\, 
\mbox{D}_{\rm{l' l}},\, \mbox{D}_{\rm{s' l}},\ 
\mbox{and}\ \mbox{D}_{\rm{l' s}}$. 
The set of observables $ \{\, \mbox{P},\, \mbox{A}_{\rm{y}},\, 
\mbox{D}_{\rm{n n}},\, \mbox{D}_{\rm{s' s}},\, \mbox{D}_{\rm{l' l}},\, 
\mbox{D}_{\rm{s' l}},\, \mbox{D}_{\rm{l' s}}\, \}$ is often referred 
to as a complete set of polarization transfer observables.
In Eq.~(\ref{e-dipj}), the symbols $\sigma_{i'}$ and $\sigma_{j}$ denote the usual
$2 \times 2$ Pauli spin matrices, and the $2 \times 2$ matrix $T$ is defined as:
\begin{eqnarray}
T\ =\ \left(
\begin{array}{cc}
T_{L J}^{s_{a} = +\frac{1}{2}, s_{a'} = +\frac{1}{2}} & 
T_{L J}^{s_{a} = -\frac{1}{2}, s_{a'} = +\frac{1}{2}} \\
T_{L J}^{s_{a} = +\frac{1}{2}, s_{a'} = -\frac{1}{2}} & 
T_{L J}^{s_{a} = -\frac{1}{2}, s_{a'} = -\frac{1}{2}} 
\end{array}\right)
\label{e-ttwobytwo}
\end{eqnarray}
where $s_{a} = \pm \frac{1}{2}$ and $s_{a'} = \pm \frac{1}{2}$ refer to the spin
projections of particles $a$ and $a'$ along the $\hat{z}$ and $\hat{z}'$ axes,
defined in Eqs.~(\ref{e-lns}) and (\ref{e-lpnsp}), respectively;
the matrix elements $T_{L J}^{s_{a}, s_{a'}}$ are related to the 
relativistic $(p,2p)$ transition
matrix element $T_{L J M_{J}}(s_{a}, s_{a'}, s_{b})$, defined in 
Eq.~(\ref{eq_6}), via
\begin{eqnarray}
T_{L J}^{s_{a}, s_{a'}}\ =\ T_{L J M_J}
(s_a, s_{a'}, s_b)\,.
\label{e-tljmrelatet}
\end{eqnarray}

\section{Results}
\label{sec:results}
\begin{figure}
\includegraphics{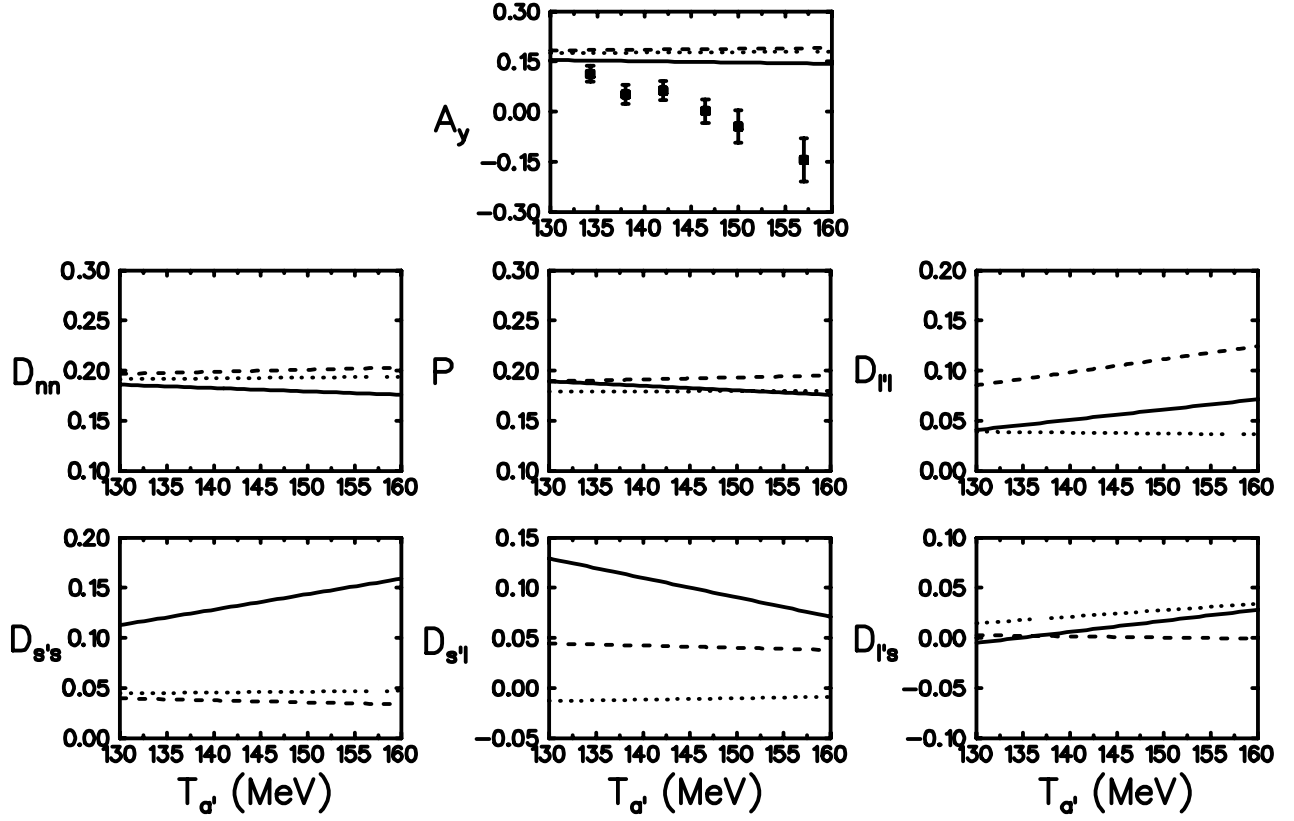}
 \caption{\label{fig_2} 
 Analyzing power A$_{\rm{y}}$, induced polarization P and spin transfer coefficients
 $\mbox{D}_{\rm{i'j}}$ plotted as a function of the laboratory kinetic 
energy $\mbox{T}_{a'}$ of the
 outgoing proton $a'$, for proton knockout from the $3s_{1/2}$ orbital of 
 $^{208}$Pb for an incident laboratory kinetic energy of 202~MeV and coplanar
 scattering angles $(\theta_{a'} = 28.0^{\circ}, \theta_{b} = -54.6^{\circ})$.
The solid line represents the IA1 calculation, the dashed line the full IA2 
calculation, and the dotted line the IA2 calculation employing 
only subclass $F^{\{+\, +\, ; \, +\, + \}}$.
The data are from Ref. \cite{Ne02}.
}
\end{figure}

\begin{figure}
\includegraphics{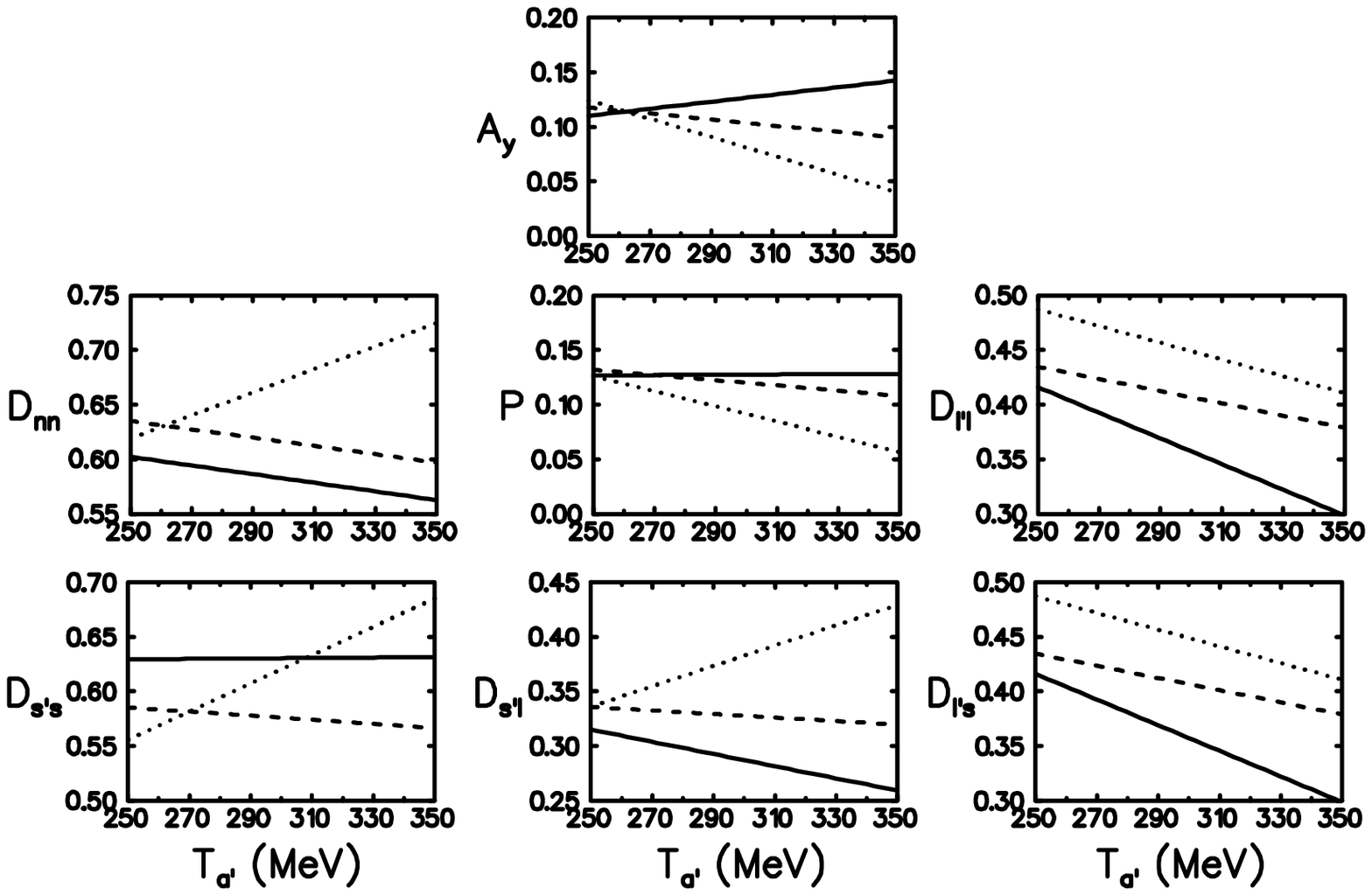}
 \caption{\label{fig_4} 
 Analyzing power A$_{\rm{y}}$, induced polarization P and spin transfer coefficients
 $\mbox{D}_{\rm{i'j}}$ plotted as a function of the laboratory kinetic energy $\mbox{T}_{a'}$ of the
 outgoing proton $a'$, for proton knockout from the $3s_{1/2}$ orbital of 
 $^{208}$Pb for an incident laboratory kinetic energy of 392~MeV and coplanar
 scattering angles $(\theta_{a'} = 32.5^{\circ}, \theta_{b} = -80.0^{\circ})$.
 The solid line represents the IA1 calculation, the dashed line the full IA2 
calculation, and the dotted line the IA2 calculation employing 
only subclass $F^{\{+\, +\, ; \, +\, + \}}$. 
}
\end{figure}

We now study the sensitivity of complete sets of polarization transfer observables
to both IA1 and IA2 representations of the NN scattering matrix. For this particular 
study we consider two kinematic conditions, namely proton knockout from the 
$3s_{1/2}$ state of $^{208}$Pb at an incident energy of 202~MeV for coplanar 
scattering angles $(\theta_{a'},\theta_{b}) \, = \, (28.0^{\circ}, -54.6^{\circ})$, 
as well as an incident energy of 392~MeV for coplanar scattering angles
$(\theta_{a'},\theta_{b}) \, = \, (32.5^{\circ}, -80.0^{\circ})$.
In general, for quantitative predictions, a spin observable can be 
regarded as being sensitive to a particular model ingredient if the inclusion 
thereof changes the observable by more than the expected maximum experimental 
error of about $\pm$ 0.1. However, since the plane wave results of this paper 
are merely qualitative, we avoid comparisons to data. Our predictions for complete 
sets of polarization transfer observables at 202 and 392~MeV are displayed in 
Figs.~(\ref{fig_2}) and (\ref{fig_4}) respectively: the analyzing power A$_{\rm{y}}$, 
induced polarization P and spin transfer coefficients
$\mbox{D}_{\rm{i'j}}$ are plotted as a function of the laboratory 
kinetic energy $\mbox{T}_{a'}$ of the outgoing proton $a'$. The solid and dashed 
lines represent calculations based on the IA1 and IA2 representations, 
respectively, whereas the dotted line represents the IA2 calculation employing 
only subclass $F^{\{+\, +\, ; \, +\, + \}}$. 
The deviation of the latter predictions from the full IA2 result (dashed line) serves as 
an indication of the importance of subclass  
$F^{\{+\, +\, ; \, -\, + \}}$ for
describing spin observables. At 202~MeV [Fig.~(\ref{fig_2})], we see 
that $\mbox{A}_{\rm{y}}$, $\mbox{D}_{\rm{l' l}}$, $\mbox{D}_{\rm{s' s}}$ and $\mbox{D}_{\rm{s' l}}$ 
all discriminate between the IA1 and IA2 representations. In contrast, P,
$\mbox{D}_{\rm{n n}}$ and $\mbox{D}_{\rm{l' s}}$ are virtually identical for both representations.
To understand these results we first expand the boundstate spinor in terms of a
Dirac plane wave basis as follows:
\begin{eqnarray}
\nonumber
 \phi_{LJ M_{J}} (\vec{K} \,)
 & = &
 \alpha_{1} U(\vec{K}, m, s_{z} = \frac{1}{2})\ +\  
 \alpha_{2} U(\vec{K}, m, s_{z} = -\frac{1}{2})\ + 
\\
 & &
 \label{eq_31}
\alpha_{3} V(\vec{K}, m, s_{z} = \frac{1}{2})\ +\  
 \alpha_{4} V(\vec{K}, m, s_{z}=  -\frac{1}{2})\,,
\end{eqnarray}
where
\begin{eqnarray}
\label{eq_32}
 \alpha_{i} & = & \alpha_{i} (L,J,M_{J},\vec{K} \, ) 
 \mbox{   for } \, i \, = \, 1, 2, 3, 4
\end{eqnarray}
and the negative-energy Dirac spinor, denoted by $V$, is given by:
\begin{eqnarray}
\label{eq_33}
 V(\vec{K}, s_{z})
 & = &
 \left[
       \frac{E(\vec{K}) + m}{2 m}
 \right]^{\frac{1}{2}} \,
 \left(
       \begin{array}{c}
	\displaystyle \frac{\vec{\sigma} \cdot \vec{K}}{E(\vec{K}) + m} \, \chi_{s_{z}}             \\
       \chi_{s_{z}}
       \end{array}
 \right).
\end{eqnarray}
The expansion coefficients $(\alpha)$ are determined from the relations:
\begin{eqnarray}
 \alpha_{1} & = & \overline{U}(\vec{K},m,s_{z} = \frac{1}{2}) \,
 \phi_{LJ M_{J}} (\vec{K} \,)\,, \nonumber\\
 \alpha_{2} & = & \overline{U}(\vec{K},m,s_{z}=-\frac{1}{2}) \,
 \phi_{LJ M_{J}} (\vec{K} \,)\,,\nonumber\\
 \alpha_{3} & = & -\overline{V}(\vec{K},m,s_{z}=\frac{1}{2}) \,
 \phi_{LJ M_{J}} (\vec{K} \,)\,,\nonumber\\
 \alpha_{4} & = & -\overline{V}(\vec{K},m,s_{z}=-\frac{1}{2}) \,
 \phi_{LJ M_{J}} (\vec{K} \,)\,,
\label{eq_39}
\end{eqnarray}
where the usual orthogonality conditions for the Dirac spinors have been used \cite{Bj64}.
The effect of the energy projection operator on the Dirac spinor can now 
clearly be identified:
\begin{eqnarray}
\label{eq_43}
 \Lambda_{\rho} (\vec{K},M) \, \phi_{LJ M_{J}} (\vec{K})
 & = &
 \left\{
   \begin{array}{c}
    \alpha_{1} U(\frac{1}{2}) + \alpha_{2} U(-\frac{1}{2}) \qquad \mbox{if  }
    \rho \, = \, +
    \\
    \alpha_{3} V(\frac{1}{2}) + \alpha_{4} V(-\frac{1}{2}) \qquad \mbox{if  }
    \rho \, = \, -
   \end{array}
 \right.
\end{eqnarray}
where we employ the shorthand notation:
\begin{eqnarray}
 U(s_z) & = & U(\vec{K},M,s_{z})\,,\nonumber\\
\label{eq_45}
 V(s_z) & = & V(\vec{K},M,s_{z})\,.
\end{eqnarray}
For the IA1 representation, substitution of Eq.~(\ref{eq_31}) into Eq.~(\ref{eq_20}) yields
\begin{eqnarray}
\nonumber
 T_{LJ M_{J}}^{IA1} (s_{a}, s_{a'}, s_{b})
 & = &
 \sum_{L=S}^{T} \, F_{L} \, 
 \left[\,
       \overline{U}_{a'} \otimes \overline{U}_{b}
 \right] \, 
 \left[
       \lambda^{L} \otimes \lambda_{L}
 \right] \,
 \left[
       \alpha_{1} U_{a} \otimes U(\frac{1}{2}) + \alpha_{2}
       U_{a} \otimes U(-\frac{1}{2}) +
 \right.
 \\
 & &
 \label{eq_46}
 \left.
       \alpha_{3} U_{a} \otimes V(\frac{1}{2}) +
       \alpha_{4} U_{a} \otimes V(-\frac{1}{2})
 \right].
\end{eqnarray}
On the other hand, for the IA2 representation, substitution of Eq.~(\ref{eq_31}) 
into Eq.~(\ref{eq_22}), and employing Eq.~(\ref{eq_43}), leads to
\begin{eqnarray}
\nonumber
\hspace{-1.0cm}T_{LJ M_{J}}^{IA2} (s_{a}, s_{a'}, s_{b})
 & = &
 \sum_{n=1}^{5} 
F_{n}^{\{+ \, + \, ; \, +  + \}} 
 \left[\,
       \overline{U}_{a'} \otimes \overline{U}_{b}
 \right] K_{n} 
 \left[
       \alpha_{1} U_{a} \otimes U(\frac{1}{2}) + 
       \alpha_{2} U_{a} \otimes U(-\frac{1}{2})
 \right] +
 \\
 & &
\label{eq_47}
\hspace{-0.5cm} 
\sum_{n=1}^{13}F_{n}^{\{+ \, + \, ; \, - \, + \}}
 \left[\,
       \overline{U}_{a'} \otimes \overline{U}_{b}
 \right] K_{n} 
 \left[
       \alpha_{3} U_{a} \otimes V(\frac{1}{2}) +
       \alpha_{4} U_{a} \otimes V(-\frac{1}{2})
 \right].
\end{eqnarray}
Note that in Eq.~(\ref{eq_47}) the first sum is only over five amplitudes, that is
$n = 1$ to 5 in 
$F_{n}^{\{+\, +\, ; \, +\, + \}}$: 
the other eight remaining amplitudes ($n = 6$ to 13) are identically zero in this subclass. 
The five non-zero amplitudes are identical to the 
amplitudes of the IA1 representation, i.e.,
\begin{eqnarray}
\label{eq_48}
\sum_{n=1}^{5} \, 
F_{n}^{\{+\, + \, ; \, + \, + \}} \,
K_{n}\ & \equiv\ & \sum_{L=S}^{T} \, F_{L} \,  
\lambda^{L} \otimes \lambda_{L}\,.
\end{eqnarray}
For free NN scattering [where the boundstate wave function, $\phi_{LJ M_{J}}(\vec{K}\,)$
in Eq.~(\ref{eq_6}) is replaced by a free positive-energy Dirac spinor], the IA1 and IA2 
representations give identical results for all spin observables. Comparison 
of Eqs.~(\ref{eq_46}) and (\ref{eq_47}) brings to light a very important difference between 
applications of the IA1 and IA2 representations to exclusive proton knockout reactions. 
For the IA1 representation we see that the 
negative-energy matrix elements [third and fourth terms in Eq.~(\ref{eq_46})]
are multiplied by positive-energy amplitudes [$F_{L}$ in Eq.~(\ref{eq_46})]. 
This is in direct contrast to the IA2 representation where
the energy projection operators ensure that positive-energy amplitudes only
couple to positive-energy matrix elements,
$F^{\{+\, +\, ; \, +\, + \}}$. 
The negative-energy matrix elements only come into play in subclass 
$F^{\{+\, +\, ; \, -\, + \}}$
in Eq.~(\ref{eq_47}). 

\begin{figure}
\includegraphics{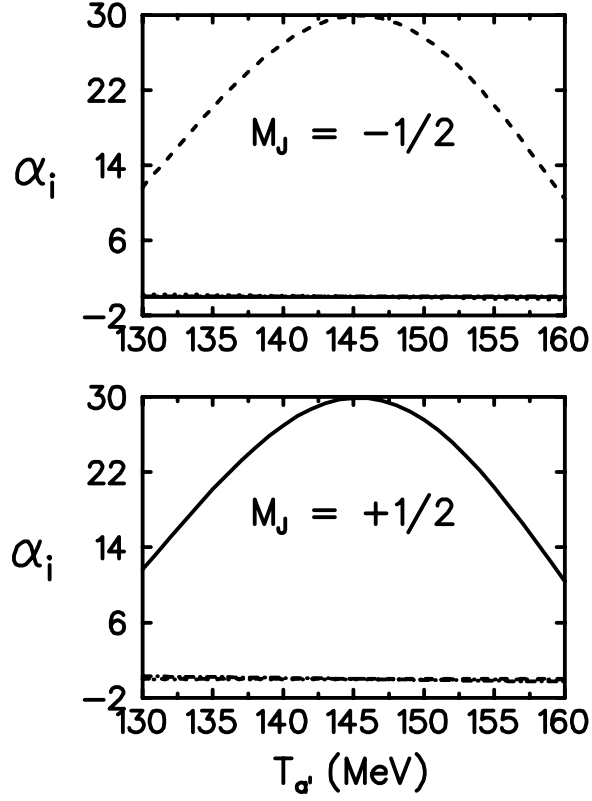}
 \caption{\label{fig_3} 
Expansion coefficients $\alpha_{1}$ (solid line), $\alpha_{2}$ (dashed line), 
$\alpha_{3}$ (dash-dotted line), and $\alpha_{4}$ (dotted line),
of the $\phi_{LJ, M_{J} = -\frac{1}{2}}(\vec{K}\,)$ [top panel]
and $\phi_{LJ, M_{J} = +\frac{1}{2}}(\vec{K})\,$ [bottom panel]
for proton knockout from the $3s_{1/2}$ state in $^{208}$Pb,
for an incident laboratory kinetic energy of 202~MeV and coplanar
scattering angles $(\theta_{a'} = 28.0^{\circ}, \theta_{b} = -54.6^{\circ})$,
in terms of free Dirac plane waves [see Eq.~(\ref{eq_31})].
}
\end{figure}

In order to understand why IA1 and IA2 predictions of some of the spin 
observables are virtually identical in Fig.~(\ref{fig_2}),
we display in Fig.~(\ref{fig_3}) the expansion coefficients $\alpha_{1}$ (solid line), 
$\alpha_{2}$ (dashed line),  $\alpha_{3}$ (dash-dotted line), and $\alpha_{4}$ (dotted line),
of the $\phi_{LJ, M_{J} = -\frac{1}{2}}(\vec{K}\,)$ [top panel] and 
$\phi_{LJ, M_{z} = +\frac{1}{2}}(\vec{K})\,$ [bottom panel] for proton knockout 
from the $3s_{1/2}$ state in $^{208}$Pb at 202~MeV, in terms of free Dirac 
plane waves [see Eq.~(\ref{eq_31})]. 
It is clearly seen that one of the positive-energy expansion 
coefficients ($\alpha_{1}$, $\alpha_{2}$) is dominant relative 
to both negative-energy expansion coefficients ($\alpha_{3}$, $\alpha_{4}$).
This implies that the negative-energy components play a negligible role 
when the spin observable displays little sensitivity to the two different
representations.

\begin{figure}
\includegraphics{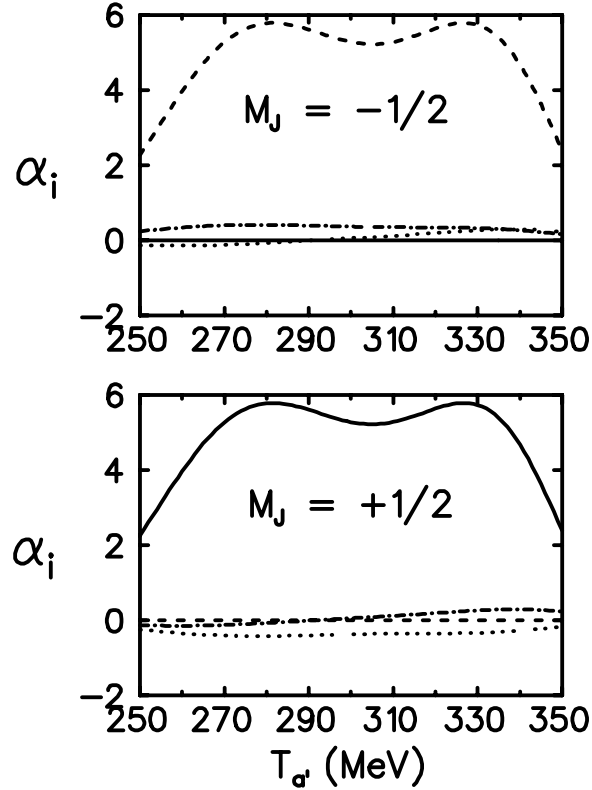}
 \caption{\label{fig_5} 
Expansion coefficients $\alpha_{1}$ (solid line), $\alpha_{2}$ (dashed line), 
$\alpha_{3}$ (dash-dotted line), and $\alpha_{4}$ (dotted line),
of the $\phi_{LJ, M_{J} = -\frac{1}{2}}(\vec{K}\,)$ [top panel] 
and $\phi_{LJ, M_{J} = +\frac{1}{2}}(\vec{K})\,$ [bottom panel]
for proton knockout from the $3s_{1/2}$ state in $^{208}$Pb,
for an incident laboratory kinetic energy of 392~MeV and coplanar
 scattering angles $(\theta_{a'} = 32.5^{\circ}, \theta_{b} = -80.0^{\circ})$,
in terms of free Dirac plane waves [see Eq.~(\ref{eq_31})].
}
\end{figure}

For exclusive proton knockout at 392~MeV the figures corresponding to Figs.~(\ref{fig_2}) and 
(\ref{fig_3}) are Figs.~(\ref{fig_4}) and (\ref{fig_5}), respectively. At this higher incident
energy we see in Fig.~(\ref{fig_4}) that most observables clearly discriminate between 
IA1 and IA2 representations, with the induced polarization P being the least sensitive. 
However, all observables are sensitive to negative energy components as can be seen by
comparing the dashed and dotted lines. It follows that at higher energies, the role of
the additional subclasses present in IA2 will increase in importance.
As in the case of the 202~MeV predictions, in Fig.~(\ref{fig_5}) we see that one of the 
positive-energy expansion coefficients ($\alpha_{1}$, $\alpha_{2}$) is dominant 
relative to both negative-energy expansion coefficients ($\alpha_{3}$, $\alpha_{4}$).  
However, contrary to the 202~MeV case, we see that percentage wise the latter
coefficients for 392~MeV are less negligible compared to the corresponding coefficients at
202~MeV, that is the ratio of positive- to negative-energy expansion coefficients.

To conclude this section we comment on the difference between using IA1 (solid line) and
IA2, but including only subclass $\hat{F}^{11}$ (dotted line) in Figs.~\ref{fig_2} and
\ref{fig_4}. For 202 MeV it is only $D_{{\rm l' l}}$ and $D_{{\rm s' s}}$ that display a
significant difference between the two calculations. At 392 MeV this difference is more
pronounced and visible in all the spin observables. Comparison of Eqs.~(\ref{eq_46}) and
(\ref{eq_47}) shows that
\begin{eqnarray}
\label{eq_49}
 T_{LJ M_{J}}^{IA1} (s_{a}, s_{a'}, s_{b})
 & = &
 T_{LJ M_{J}}^{IA2,11} (s_{a}, s_{a'}, s_{b}) + \Delta
\end{eqnarray}
where $T_{LJ M_{J}}^{IA2,11} (s_{a}, s_{a'}, s_{b}) $ means the inclusion of only
subclass $\hat{F}^{11}$ and 
\begin{eqnarray}
\label{eq_50}
\Delta & = &
 \sum_{L=S}^{T} \, F_{L} \, 
 \left[\,
       \overline{U}_{a'} \otimes \overline{U}_{b}
 \right] \, 
 \left[
       \alpha_{3} U_{a} \otimes V(\frac{1}{2}) +
       \alpha_{4} U_{a} \otimes V(-\frac{1}{2})
 \right].
\end{eqnarray}
The only difference between these two calculations therefore lies in the coupling of
positive to negative energy spinors. The quantity $\Delta$ depends on the spin
orientation of the incident and outgoing particles. However, since the spin observables
are complicated combinations of $T_{LJ M_{J}}$ it is very difficult to predict 
before-hand the degree of sensitivity to the difference between using IA1 and IA2
with only subclass $\hat{F}^{11}$. Eventhough, we have shown in Figs.~\ref{fig_3} and
\ref{fig_5} that $\alpha_{3}$ and $\alpha_{4}$ are in general small compared to 
$\alpha_{1}$ and $\alpha_{2}$, it follows from Eq.~(\ref{eq_50}) that the amplitudes
together with the matrix elements combine constructively to enhance the effect of 
$\Delta$ to $T_{LJM_{J}}$. 

\section{Summary and conclusions}
\label{sec:conclusions}
In this paper, we have studied the sensitivity of complete sets of exclusive 
($\vec{p},2 \vec{p}\,$) polarization transfer observables to different Lorentz 
invariant representations of the NN scattering matrix, namely an ambiguous 
five-term parametrization (called the IA1 representation) and an unambiguous 
and complete representation in terms of 44 invariant amplitudes (referred to 
as the IA2 representation). To avoid complications associated with the 
distortion of the scattering wave functions by the nuclear medium, the scattering
process is described within the framework of the relativistic plane wave impulse 
approximation, where the effect  of the nuclear medium on the scattering wave 
functions is neglected. 

For this particular study, we have considered two kinematic conditions, namely 
proton knockout from the $3s_{1/2}$ state of $^{208}$Pb at an incident energy of 
202~MeV for coplanar scattering angles 
$(\theta_{a'},\theta_{b}) \, = \, (28.0^{\circ}, -54.6^{\circ})$, 
as well as an incident energy of 392~MeV for coplanar scattering angles
$(\theta_{a'},\theta_{b}) \, = \, (32.5^{\circ}, -80.0^{\circ})$.
It is seen that both IA1- and IA2-based predictions give virtually identical 
results for some spin observables at 202~MeV, whereas most predictions at 392~MeV clearly
discriminate between both representations. The fact that even at the plane wave 
level, different representations predict different observables, suggests that 
one can also expect differences for the more realistic case where plane waves are 
replaced by distorted waves. Consequently, since current relativistic distorted wave 
models are based on the ambiguous IA1 parametrization, one needs to re-interpret 
all exclusive ($\vec{p},2 \vec{p}\,$) data within the framework of the relativistic
distorted wave impulse approximation based on the IA2 representation of the
NN scattering matrix. This will form the subject of a future paper.
\begin{acknowledgements}
G.C.H acknowledges financial support from the Japanese Ministry of Education, 
Science and Technology for research conducted at the Research Center for Nuclear 
Physics, Osaka, Japan. B.I.S.v.d.V gratefully acknowledges the financial support 
of the National Research Foundation of South Africa. This material is based upon 
work supported by the National Research Foundation under Grant numbers: 
GUN 2053786 (G.C.H), 2048567 (B.I.S.v.d.V).
\end{acknowledgements}

\end{document}